\documentclass[twocolumn,pra,showpacs,floatfix]{revtex4}
\usepackage{graphicx}
\usepackage{bm}
\usepackage{amsmath,amssymb,epsfig}

\newcommand{\mathi}{\mathrm{i}}
\newcommand{\mathe}{\mathrm{e}}

\newcommand{\tmtextbf}[1]{{\bfseries{#1}}}
\newcommand{\tmtextit}[1]{{\itshape{#1}}}

\newcommand{\tmem}[1]{{\em #1\/}}

\newcommand{\tmop}[1]{\ensuremath{\text{#1}}}

\begin{document}

\title{Entangling the free motion of a particle pair: an experimental scenario}

\author{Clemens Gneiting}
\author{Klaus Hornberger}
\affiliation{Arnold Sommerfeld Center for Theoretical Physics, 
Ludwig-Maximilians-Universit{\"a}t M{\"u}nchen, Theresienstra{\ss}e 37, 
80333 Munich, Germany
}

\date{\today}

\begin{abstract}
The concept of dissociation-time entanglement provides a means of
manifesting non-classical correlations in the motional state of two
counter-propagating atoms. In this article, we discuss in detail the
requirements for a specific experimental implementation, which is based
on the Feshbach dissociation of a molecular Bose-Einstein condensate of
fermionic lithium.  A sequence of two magnetic field pulses serves to
delocalize both of the dissociation products into a superposition of
consecutive wave packets, which are separated by a macroscopic distance.
This allows to address them separately in a switched Mach-Zehnder
configuration, permitting to conduct a Bell experiment with simple
position measurements. We analyze the expected form of the two-particle
wave function in a concrete experimental setup that uses lasers as atom
guides. Assuming viable experimental parameters the setup is shown to be
capable of violating a Bell inequality.
\end{abstract}


\pacs{03.67.Bg, 37.25.+k, 03.65.Ud  \hspace{\fill} \textsf{published
in Opt.~Spectroscop.~{108}, 188 (2010)}}
\maketitle

\section{\label{Introduction}Introduction}

The increasing control of quantum systems that has been obtained in the last
decades brings into reach the possibility of performing hitherto unprecedented
tests of the principles of quantum mechanics
{\cite{Sherson2006a,Gleyzes2007a,Halder2007a,Moehring2007a,Ursin2007a}}. In
particular, there is a growing interest in demonstrating non-classical
behavior in systems or with respect to features that would be assigned to the
classical realm according to our every-day experience. One exciting approach
aims at luring mesoscopic mechanical oscillators into the quantum regime, see
{\cite{Aspelmeyer2008a}} and refs therein. Another approach focuses on the
generation of highly non-classical states of continuous variables with direct
classical correspondence, such as creating superpositions of macroscopically
different phase space coordinates
{\cite{Schrodinger1935a,Monroe1996a,Raimond2001a,Gerlich2007a}}. The recent
achievements in the generation and manipulation of ultracold atoms and
molecules
{\cite{Donley2002a,Jochim2003a,Mukaiyama2004a,Durr2004b,Kohler2006a,Jones2006a}}
suggest to imprint non-classical features even on the free motion of material
particles. For example, it has been proposed to perform the original
Einstein-Podolsky-Rosen thought experiment based on the dissociation of a
molecular Bose-Einstein condensate (BEC)
{\cite{Einstein1935a,Kheruntsyan2005a}}.

\begin{figure*}[tb]
  \resizebox{12cm}{!}{\epsfig{file=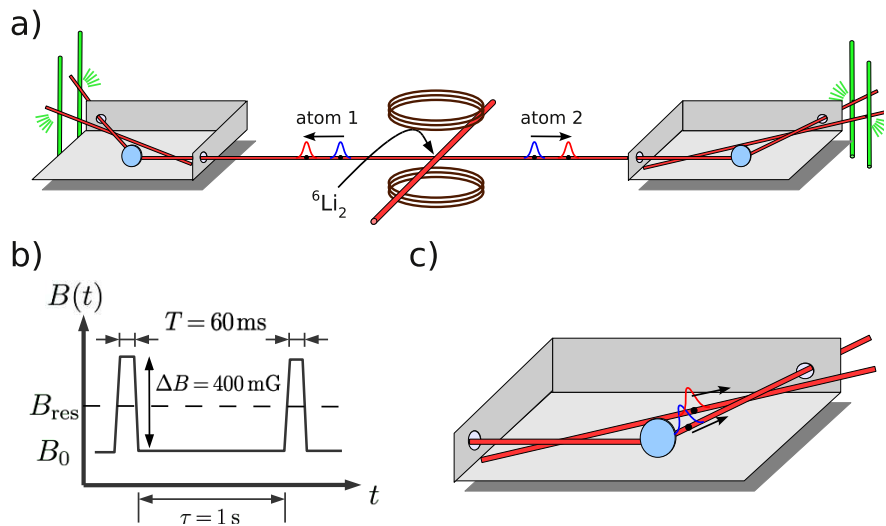}}
  \caption{\label{ExperimentalSetup}(color online) a) Experimental setup for
  the DTE Bell test. It starts with a $^6 \tmop{Li}_2$ Bose-Einstein
  condensate of about $10^2$ molecules in an optical dipole trap. A sequence
  of two short magnetic field pulses, as shown in b), dissociates on average
  one molecule per shot into a pair of counter-propagating atoms. The
  resulting two-particle state corresponds to a superposition of two
  subsequent dissociations, where each atom is delocalized into two
  consecutive wave packets that are separated by a distance of $5 \tmop{mm}$.
  The two atoms propagate in opposite directions along the laser guide and
  eventually enter switched, asymmetric Mach-Zehnder interferometers that are
  implemented by two more laser beams crossing the guide in a triangular
  arrangement at small angles. These second lasers are switched on after the
  early wave packets have passed, but before the late wave packets arrive. An
  additional perpendicular, blue-detuned blocking laser beam, which is not
  depicted in the figure, may be used to support the deflection of the late
  wave packets. This way, the early wave packets have to take the long paths,
  whereas the late ones are directed into the short paths, as depicted in c).
  If the path length differences cancel the distance between the early and the
  late wave packets, they interfere and the joint probability of detection \
  in the output ports of the interferometers exhibits a fringe pattern as a
  function of the path length difference variation. Correlations of the
  single-particle detection events thus have the potential to violate a Bell
  inequality.}
\end{figure*}

In a recent article we go one step further by proposing to violate a Bell
inequality in the motional state of material particles using position
measurements in the end {\cite{Gneiting2008a}}. The two-particle state
considered in this proposal can be characterized by its generation procedure.
Starting from a diatomic Feshbach molecule, which is trapped in an optical
dipole trap, a sequence of two short magnetic field pulses creates the
superposition of two subsequent dissociations. Due to momentum conservation
the dissociated atoms propagate in opposite directions along some wave guide,
with each atom delocalized into a pair of macroscopically separated,
consecutive wave packets, corresponding to the early and the late dissociation
instant, see Fig. \ref{ExperimentalSetup}. An observer cannot tell whether the
dissociation took place at the early or at the late instant, while both atoms
must have started their journey at the same time. In this sense, the generated
two-particle state may be called dissociation-time entangled (DTE).

Since each atom in a DTE state is described by two spatially distinct
propagating wave packets, which correspond to the early and the late
dissociation times, the use of switched, asymmetric Mach-Zehnder
interferometers can serve first to endow these wave packets with additional
relative phases and then to rejoin them. In detail, the early wave packets
propagate through the long arms of the interferometers, whereas the late ones
are directed through the short arms, see Fig. \ref{ExperimentalSetup} c). If
the detour of the early wave packets is chosen appropriately, they rejoin the
late ones in the final beam splitters. Provided the two dissociation processes
are coherent in the sense that no information about the dissociation time has
leaked to the environment, the rejoined wave packets may interfere. The joint
probability for detecting the particles in the output ports of the two
interferometers then exhibits a fringe pattern under variation of either path
length difference, which is in principle capable of violating a Bell
inequality. We emphasize that the only relevant information required to
extract these correlations is the output port combination where the atoms are
detected. This information can be obtained by simple position measurements,
without the need of a particular spatial or temporal resolution. And since the
position measurements can be understood from a classical physicist's point of
view, they are indeed in line with the above mentioned pursuit of focusing on
features that have a classical analogoue.

We note that a similar Bell test procedure was first introduced for photons in
{\cite{Brendel1999a,Tittel2000a}}. This led to the experimental demonstration
of a Bell inequality violation for parties separated by several kilometers,
albeit using a postselection procedure required due to the lack of
sufficiently good switches for photons. The implications for material
particles, in particular the detrimental effect of dispersion, are discussed
in detail in {\cite{Gneiting2008b}}.

In this article we devise a concrete scenario to perform the described DTE
Bell test, which is sufficiently detailed and realistic to suggest that
violating a Bell inequality should be possible within reach of present day
technology. Section \ref{ExperimentalScenario} provides a detailed
quantitative account of the proposed setup, based on realistic choices for the
experimental parameters. It involves pulse separations on the order of seconds
and implies atom velocities on the order of centimeters per second, as well as
a truly macroscopic spatial separation of the single atom wave packets on the
order of centimeters. In Section \ref{DTEStateGeneration} we discuss and
numerically evaluate the two-particle superposition state that is generated in
the wave guide in the course of a double dissociation with two square pulses.
Section \ref{DTEBellTest} demonstrates that this state, if exposed to the
switched, asymmetric Mach-Zehnder processing, generates a fringe pattern which
violates a Bell inequality. Finally, we conclude in Section \ref{Conclusions}
by pointing out the advantages of the proposed scheme.

\section{\label{ExperimentalScenario}Experimental scenario for a DTE Bell
test}

In the following we discuss quantitative details of a possible experimental
setup that should be capable of generating a pair of dissociation-time
entangled atoms and to violate a Bell inequality in terms of single-particle
interferometry and simple position measurements. We work at this example to
prove the viability of the proposal presented in {\cite{Gneiting2008a}}, but
of course many alternative implementations are conceivable. We emphasize that
the setup is designed to allow for time and length scales on the order of
seconds and centimeters, establishing entanglement with respect to
single-particle properties that are truly macroscopically distinct. In this
section we focus on a consistent set of experimental parameters, deferring the
theoretical analysis of the parts concerning the dissociation process to
Section \ref{DTEStateGeneration}.

We suggest to use a dilute molecular Bose-Einstein condensate (BEC) produced
from a 50:50 spin mixture of fermionic $^6 \tmop{Li}$ as a starting point.
Such a fermionic mixture seems favorable compared to bosonic ingredients,
since huge lifetimes of more than $10 \text{s}$ can be achieved due to Pauli
blocking of detrimental 3-body collisions {\cite{Jochim2003a}}. This makes
truly macroscopic time separations $\tau$ between the two dissociation pulses
conceivable, and we choose $\tau = 1 \text{s}$ in the following. It has been
demonstrated that the molecular $^6 \tmop{Li}$ BEC can be prepared efficiently
and with near-perfect purity {\cite{Jochim2003a}}, and the comparatively small
mass of lithium reconciles reasonable propagation velocities, on the order of
$1 \tmop{cm} / \text{s}$, with resolvable de Broglie wave lengths, on the
order of $10 \mu \text{m}$.

The BEC is prepared in an optical dipole trap created by two red-detuned, far
off-resonant, crossing laser beams with a wave length of, say $\lambda = 1 \mu
\text{m}$, see Fig. \ref{ExperimentalSetup} a). The guiding laser, which
connects the two single-atom interferometers, serves as an atom guide which
compensates the gravitational force. Moreover, it must be chosen sufficiently
strong that the transverse motion of the atoms remains frozen in the ground
state. This way an effectively one-dimensional description of the atomic
motion can be used.

The following numerical values are chosen to be consistent with a pulse
separation of $\tau = 1 \text{s}$ and a relative velocity of $v_{\tmop{rel}} =
1 \tmop{cm} / \text{s}$. We assume that the guiding laser is given by a
Gaussian laser beam and operated at a power of $P_{\text{G}} = 32.85 \text{W}$
with a (center) waist of $w_{0, \text{G}} = 216 \mu \text{m}$, such that the
optical dipole potential resulting from a molecular polarizability of
$\alpha_{\tmop{Li}_2} (\lambda = 1 \mu \text{m}) = 88.5 \text{{\AA}}^3$ has a
(center) trap depth of $30 \mu \text{K}$. It thus is strong enough to
compensate gravitation, supporting several thousand transversally bound
states. The implied Rayleigh length of about $z_{0, \text{G}} = 15 \tmop{cm}$
sets the scale for the extension of the whole setup. Decoherence due to photon
scattering can be estimated to occur at a rate on the order of $0.05
\text{s}^{- 1}$, which limits the overall duration of a single experiment to
remain below 20 s. The implied transverse trap frequency of $\omega_{\text{G}}
/ 2 \pi = 300 \tmop{Hz}$, on the other hand, limits the maximum dissociation
velocity not to exceed about $v_{\tmop{rel}} = 1 \tmop{cm} / \text{s}$, if we
demand the transverse motion of the atoms to remain frozen in the ground
state. The corresponding de Broglie wave length $\lambda_{\tmop{rel}} = 12.4
\mu \text{m}$, which sets the scale for the fringe separation, is in
compliance with viable stability requirements for the interferometers.

The second laser beam intersects the guiding laser perpendicularly, creating
an elongated dipole trap for the BEC within the laser guide. At the same time,
it must be sufficiently shallow to be overcome with reasonable magnetic field
pulses. Moreover, the resulting longitudinal trap ground state $|
\psi_{\text{T}} \rangle$, which is supposed to determine the center of mass
state of the trapped molecules, must meet stringent conditions regarding its
momentum spectrum. Only a sufficiently narrow momentum spectrum guarantees
that the dispersion-induced reduction of the non-local correlations due to the
momentum spread in the center of mass motion admits the violation of a Bell
inequality {\cite{Gneiting2008a}}. This demands a comparatively small trap
frequency. Specifically, for an atomic propagation velocity of $5 \tmop{mm} /
\text{s}$, the trap frequency should not greatly exceed $\omega_{\text{T}} / 2
\pi = 0.25 \tmop{Hz}$.

A reasonable trap depth of $U_{\text{T}} / k_{\text{B}} = 50 \tmop{nK}$
combined with the required trap frequency of $\omega_{\text{T}} / 2 \pi = 0.25
\tmop{Hz}$ implies the trap laser (center) beam waist to be $w_{0, \text{T}} =
1.1 \tmop{cm}$. By employing an elliptic Gaussian laser beam with waist ratio
$w_{x, \text{T}} / w_{y, \text{T}} = 10$, the required laser power can be kept
reasonably low, at $P_{\text{T}} = 13.1 \text{W}$. The corresponding photon
scattering rate of $10^{- 4}  \text{s}^{- 1}$, on the other hand, does not
impose any limitations.

The preparation is arranged such that only a small number of molecules, on the
order of $10^2$, remains in the BEC at the end. These can be taken to be
non-interacting, so that the initial longitudinal center of mass state of the
molecules indeed is given by the trap laser ground state, parametrized by the
trap frequency $\omega_{\text{T}}$.

A DTE atom pair is extracted from the BEC by applying a sequence of two short
square pulses in the external magnetic field, as shown in Fig.
\ref{ExperimentalSetup} b). The pulses must be chosen such that the desired
dissociation velocity of $v_{\tmop{rel}} = 1 \tmop{cm} / \text{s}$ is
accomplished. To this end the pulses not only have to provide the
corresponding kinetic energy but also the energy to overcome the longitudinal
trap depth of $U_{\text{T}} = k_{\text{B}} 50 \tmop{nK}$. Moreover, the
generated state of the relative motion must meet similarly stringent
conditions concerning its momentum spectrum as the trap laser ground state $|
\psi_{\text{T}} \rangle$.

The desired dissociation velocity of $v_{\tmop{rel}} = 1 \tmop{cm} / \text{s}$
is obtained for a narrow, sufficiently isolated resonance with, e.g.,
resonance width $\Delta B_{\tmop{res}} = 1 \tmop{mG}$ and $\mu_{\tmop{res}} =
0.01 \mu_{\text{B}}$, by applying a pulse duration of $T = 60 \tmop{ms}$ and a
pulse height of $B_0 + \Delta B - B_{\tmop{res}} = 200 \tmop{mG}$. Here,
$\mu_{\tmop{res}}$ denotes the difference between the magnetic moments of the
resonance state and the open channel, $B_0$ stands for the magnetic field base
value before and after the pulses, and $B_{\tmop{res}}$ indicates the
resonance position. In particular, the generated state meets the tight
restrictions on the momentum distribution as imposed by dispersion. In Section
\ref{DTEStateGeneration} we will analyze the momentum spectrum resulting from
such a double pulse sequence and explicitly demonstrate its capability of
violating a Bell inequality.

The dissociation probability per double pulse sequence depends, apart from the
shape of the magnetic field pulses, on the resonance parameters $\Delta
B_{\tmop{res}}$ and $\mu_{\tmop{res}}$, on the guiding laser trapping
frequency $\omega_{\text{G}}$, and on the background channel scattering length
$a_{\tmop{bg}}$. For a generic value of $a_{\tmop{bg}} = 100 a_0$ the
dissociation probability amounts to a few percent. We thus rely on
post-selection in order to guarantee that only one atom pair per shot enters
the interferometers. This is however unproblematic, since the final
fluorescence detection of the slow, strongly confined atoms can be done with
single particle resolution, so that it is easy to disregard the cases of too
many atom pairs in the process. For a start, we stick to this simple
post-selection procedure; in a more refined setup it is conceivable to use a
specially prepared optical lattice where each site is occupied by at most one
molecule {\cite{Volz2006a}}.

After the completion of the dissociation sequence, and for our chosen time
separation of $\tau = 1 \text{s}$ between the early and the late dissociation
pulse, the corresponding early and late wave packets of each particle
propagate at a velocity of 5 mm/s, separated by a distance of 5 mm on each
side. This constitutes a truly macroscopic delocalization of each atom.
Immediately after the dissociation process, the widths of the early and late
single-particle wave packets are on the order of about $200 \mu \text{m}$.
Their narrow momentum spectra guarantee that these wave packet extensions are
not appreciably modified during the propagation to the interferometers if the
propagation time does not exceed about $10 \text{s}$. The early and late wave
packets are thus spatially still sufficiently distinct when arriving at the
interferometers for the switches to be applicable.

The interferometers are implemented by two more red-detuned laser beams
crossing the guide in a triangular arrangement at small angles, see Fig.
\ref{ExperimentalSetup}. The path length differences must compensate the $5
\tmop{mm}$ distance between the early and late wave packets of each particle.
Small variations of the path length differences on the order of the de Broglie
wave length $\lambda_{\tmop{rel}} = 12.4 \mu \text{m}$ result in a variation
of the relative phase between the early and the late wave packets, as required
for the Bell test {\cite{Gneiting2008a}}. The additional laser beams are
switched on only after the early wave packets have passed the crossing points,
but before the late ones arrive there. The crossings of the beams then act as
beam splitters, while the required atom mirrors may be realized using
evanescent light fields or blue-detuned laser beams perpendicular to the
interferometer planes {\cite{Adams1994a,Kreutzmann2004a}}. Such perpendicular
blocking beams at the crossing point of the first beam splitters could also
support the full deflection of the late wave packets into the short arms. This
way, the early wave packets pass through the long arms, whereas the late wave
packets are deflected into the short arms, as depicted in Fig.
\ref{ExperimentalSetup} c). Note that a simplified setup could do without the
switching, replacing the switches by ordinary beam splitters, at the cost of
50\% post-selection.

Resonant laser beams crossing the guiding lasers behind the final beam
splitters, as sketched in Fig. \ref{ExperimentalSetup} a), effect fluorescence
detection with near unit efficiency and with single particle resolution. No
particular spatial or temporal resolution of the detection is required, since
the only relevant information for the Bell test is the combination of beam
splitter output ports where the particles are detected. Similar to the
original Bell test based on spins, the joint probability distribution for
detecting the particles in particular output port combinations shows an
interference pattern under variation of either path length difference, which
can violate a Bell inequality. Decoherence due to photon scattering in the
guiding laser beam sets the scale for the distance between the BEC trap and
the detection lasers to be on the order of 10 cm in our example.

\section{\label{DTEStateGeneration}DTE state from a double dissociation pulse}

We proceed to derive the asymptotic form of the two-particle state that is
generated by a double pulse sequence in the described setup, and we verify
that it has the DTE structure as described in Section \ref{Introduction}. In
particular, we will be able to specify the corresponding momentum spectrum,
which plays a crucial role for the feasibility of the Bell test. Below in
Section \ref{DTEBellTest} we will use this result to demonstrate explicitly
that the experimental scenario presented above yields a DTE state which can
violate a Bell inequality in spite of wave packet dispersion.

A two-channel single-resonance calculation shows that after an arbitrary
magnetic field pulse sequence (close to an isolated resonance) the dissociated
part of the state, $| \Phi_{\tmop{bg}} (t) \rangle$, is described, for low
energies, at positions far from the dissociation center, and at large times,
by the asymptotic form
\[ | \Phi_{\tmop{bg}} (t) \rangle \sim C_{\tmop{bg}} | \varphi^{\tmop{cm}}_{0,
   0} \rangle | \varphi^{\tmop{rel}}_{0, 0} \rangle \widehat{\text{U}}_{z,
   t}^{(0)} | \Psi_z \rangle . \]
Here $\widehat{\text{U}}_{z, t}^{(0)}$ denotes the free (two-particle)
propagator in the longitudinal direction, while the transverse motion is
frozen in the harmonic ground state, $| \varphi^{\tmop{cm}}_{0, 0} \rangle$
and $| \varphi^{\tmop{rel}}_{0, 0} \rangle$, resp., of the guiding laser beam.
The longitudinal part of the two-particle state, expressed in the basis of the
center of mass and relative momenta $p_{\tmop{cm}}$ and $p_{\tmop{rel}}$, is
determined by {\cite{Gneiting2008a}}
\begin{equation}
  \label{MomentumSpectrum} \langle p_{\tmop{cm}}, p_{\tmop{rel}} | \Psi_z
  \rangle = \frac{\tilde{C} (p^2_{\tmop{cm}} / 4 m \hbar + p^2_{\tmop{rel}} /
  m \hbar + 2 \omega_{\text{G}})}{\| \tilde{C} \|} \langle p_{\tmop{cm}} |
  \psi_{\text{T}} \rangle,
\end{equation}
where $\tilde{C} (\omega)$ is the Fourier transform of the closed channel
probability amplitude $C (t)$, $\tilde{C} (\omega) = \int^{\infty}_{- \infty}
\text{d} t \mathe^{\mathi \omega t} C (t)$. The quantity
\begin{eqnarray}
  \label{SpectrumNorm} \| \tilde{C} \|^2 &=& \int^{\infty}_{- \infty} \text{d}
  p_{\tmop{cm}}  \int^{\infty}_{- \infty} \text{d} p_{\tmop{rel}} \left| \tilde{C}
  \left( \frac{p^2_{\tmop{cm}}}{4 m \hbar} + \frac{p^2_{\tmop{rel}}}{m \hbar}
  + 2 \omega_{\text{G}} \right) \right|^2 
\nonumber\\
&&\times| \langle p_{\tmop{cm}} | \psi_{\text{T}}
  \rangle |^2
\end{eqnarray}
normalizes the spectrum. The dissociation probability can be estimated by
\begin{equation}
  \label{DissociationProbability} |C_{\tmop{bg}} |^2 = \frac{\omega_{\text{G}}
  a_{\tmop{bg}} \mu_{\tmop{res}} \Delta B_{\tmop{res}}}{\pi \hbar^2} \|
  \tilde{C} \|^2,
\end{equation}
which involves the background scattering length $a_{\tmop{bg}}$, the resonance
width $\Delta B_{\tmop{res}}$, and $\mu_{\tmop{res}}$, the difference between
the magnetic moments of the resonance state and the open channel. As mentioned
in Section \ref{ExperimentalScenario}, we presume a narrow resonance width of
$\Delta B_{\tmop{res}} = 1 \tmop{mG}$ and short pulse durations $T = 60
\tmop{ms}$, such that only a small fraction of the condensate dissociates. The
coupling between the two channels is then sufficiently weak so that the
back-action from the open channel to the bare resonance state can be safely
neglected. In this case, a time-dependent magnetic field $B (t)$ determines
the closed channel probability amplitude according to
\begin{equation}
  \label{ClosedChannelAmplitude} C (t) = C (t_0) \exp \left( -
  \frac{\mathi}{\hbar}  \int^t_{t_0} \text{d} t' [E_{\tmop{res}} (B (t')) - 2
  U_{\text{T}} + \hbar \omega_{\text{G}}] \right) .
\end{equation}
The resonance state energy $E_{\tmop{res}} (B (t)) = \mu_{\tmop{res}} (B (t)
- B_{\tmop{res}})$ is measured with respect to the background channel
continuum threshold, whereas the trap laser potential depth $U_{\text{T}}$ and
the transverse trap frequency of the guiding laser $\omega_{\text{G}}$ cause
an additional energy offset that has to be overcome in the dissociation
process. $C (t_0)$ indicates the probability amplitude for the two particles
to be found in the resonance state at some initial time $t_0$ well before the
dissociation pulses take place. Since we can presume that the magnetic field
before $t_0$ has remained well below the resonance $B_{\tmop{res}}$, the
corresponding background channel component is negligible and we can safely
set $C (t_0) = 1$ from now on.

\begin{figure*}[tb]
  \resizebox{14cm}{!}{\epsfig{file=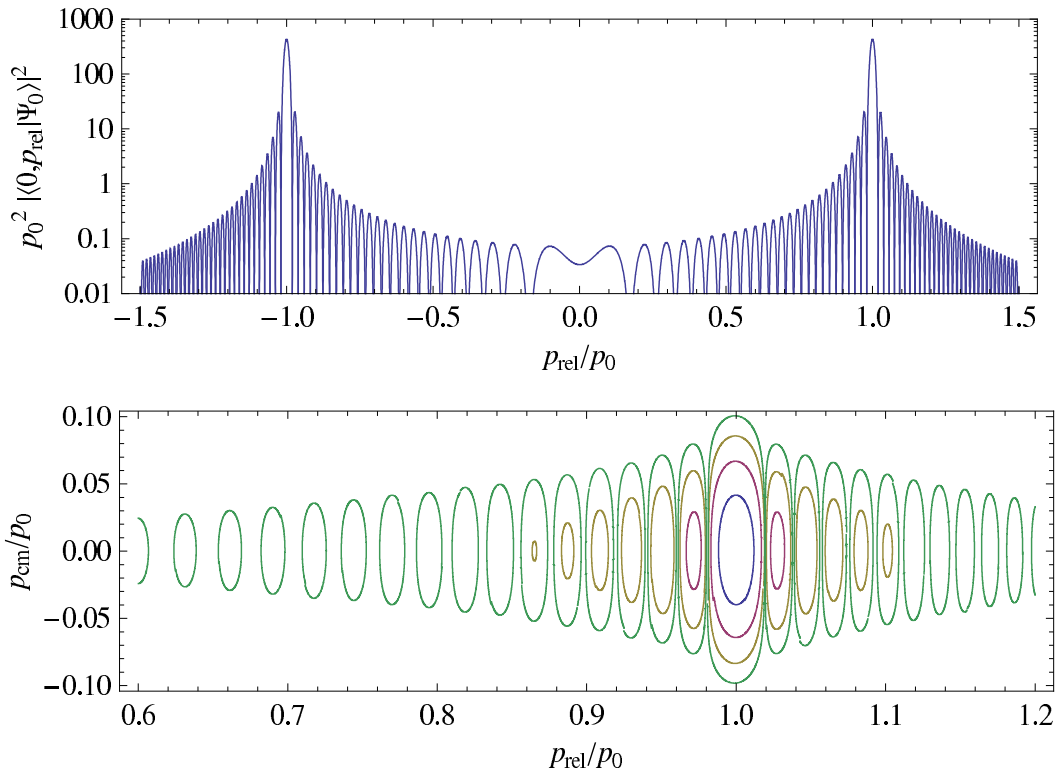}}
  \caption{\label{SinglePulseSpectrumPlot}(color online) Momentum distribution
  $| \langle p_{\tmop{cm}}, p_{\tmop{rel}} | \Psi_0 \rangle |^2$ for a single
  Feshbach dissociation pulse with pulse duration $T = 60 \tmop{ms}$, magnetic
  field base value $B_{\tmop{res}} - B_0 = 200 \tmop{mG}$ and pulse height
  $\Delta B = 400 \tmop{mG}$. Assuming a longitudinal trap depth of
  $U_{\text{T}} / k_B = 50 \tmop{nK}$ and a transverse trapping frequency of
  $\omega_{\text{G}} / 2 \pi = 300 \tmop{Hz}$ this yields the dissociation
  velocity $p_0 / m = 5 \tmop{mm} / \text{s}$. The upper plot shows the
  momentum distribution in the $p_{\tmop{cm}} = 0$ plane on a logarithmic
  scale. As required, the distribution exhibits sharp peaks at
  $(p_{\tmop{cm}}, p_{\tmop{rel}}) = (0, \pm p_0)$, corresponding to
  symmetrically counterpropagating particles. The lower logarithmic contour
  plot focuses on the vicinity of the $(p_{\tmop{cm}}, p_{\tmop{rel}}) = (0, +
  p_0)$ peak, including the dependence on the center of mass momentum
  $p_{\tmop{cm}}$. The width in $p_{\tmop{rel}}$ is characterized by $\Delta
  p^2 = 2 m \hbar / T$, with $(\Delta p / p_0)^2 = 0.012$, whereas the width
  in $p_{\tmop{cm}}$ is essentially determined by the momentum uncertainty
  $\sigma_{p, \text{T}}$ of the longitudinal trap laser ground state $|
  \psi_{\text{T}} \rangle$, which follows from the trap frequency
  $\omega_{\text{T}} / 2 \pi = 0.25 \tmop{Hz}$, yielding $\sigma_{p, \text{T}}
  / p_0 = 0.024$.}
\end{figure*}

\begin{widetext}
For a sequence of two square pulses, each of height $\Delta B$ and duration
$T$, separated by the time $\tau$,
\begin{eqnarray} B (t) &=& B_0 + \Delta B \theta \left( t + \tau + \frac{T}{2} \right) \theta
   \left( - \tau + \frac{T}{2} - t \right) 
+ \Delta B \theta \left( t +
   \frac{T}{2} \right) \theta \left( \frac{T}{2} - t \right), 
\end{eqnarray}
the Fourier transform of (\ref{ClosedChannelAmplitude}) can be evaluated
analytically, yielding
\begin{eqnarray}
  \tilde{C} (\omega) & = & \frac{T \mu_{\tmop{res}} \Delta B \tmop{sinc}
  \left[ \left( \omega - \mu_{\tmop{res}} [B_0 + \Delta B - B_{\tmop{res}}] /
  \hbar + 2 U_{\text{T}} / \hbar - \omega_{\text{G}} \right) T / 2
  \right]}{\hbar \omega - \mu_{\tmop{res}} [B_0 - B_{\tmop{res}}] + 2
  U_{\text{T}} - \hbar \omega_{\text{G}}}  \label{DoublePulseSpectrum}\\
  &  & \times \left\{ \mathe^{- \mathi \omega \tau} + \mathe^{\mathi [2
  U_{\text{T}} \tau - \mu_{\tmop{res}} \Delta B T + \mu_{\tmop{res}}
  (B_{\tmop{res}} - B_0) \tau - \hbar \omega_{\text{G}} \tau] / \hbar}
  \right\}, \nonumber
\end{eqnarray}
\end{widetext}
where we have neglected an irrelevant global phase. It is now helpful to
introduce the mean energy
\[ \frac{p^2_0}{m} = \mu_{\tmop{res}} (B_0 + \Delta B - B_{\tmop{res}}) - 2
   U_{\text{T}} - \hbar \omega_{\text{G}}, \]
the characteristic width
\[ \Delta p^2 = 2 m \hbar / T, \]
and the pulse energy
\[ \bar{p}^2 / m = \mu_{\tmop{res}} \Delta B. \]
The dissociated state (\ref{MomentumSpectrum}) thus takes the form
\begin{eqnarray}
  \label{NonNormalizedSpectrum}  
  \| \tilde{C} \| \langle p_{\tmop{cm}}, p_{\tmop{rel}} | \Psi_z \rangle & =& T \bar{p}^2
  \langle p_{\tmop{cm}} | \psi_{\text{T}} \rangle
\\
&&\times  \frac{
  \tmop{sinc} [ \left( p^2_{\tmop{cm}} / 4 + p^2_{\tmop{rel}} - p^2_0 \right)
  / \Delta p^2]}{p^2_{\tmop{cm}} / 4 + p^2_{\tmop{rel}} - p^2_0 + \bar{p}^2} 
\nonumber\\
  &  & \times \left\{ \mathe^{- \mathi (p^2_{\tmop{cm}} / 4 m \hbar +
  p^2_{\tmop{rel}} / m \hbar) \tau} + \mathe^{\mathi \phi_{\tau}} \right\},
\nonumber
\end{eqnarray}
where the relative phase between the early and the late dissociation component
is given by
\[ \phi_{\tau} = [2 U_{\text{T}} \tau - \mu_{\tmop{res}} \Delta B T +
   \mu_{\tmop{res}} (B_{\tmop{res}} - B_0) \tau] / \hbar + \omega_{\text{G}}
   \tau . \]
For the choice of parameters presented in Section \ref{ExperimentalScenario}
the momentum spectrum (\ref{NonNormalizedSpectrum}) is sharply peaked at
$(p_{\tmop{cm}}, p_{\tmop{rel}}) = (0, \pm p_0)$, with $p_0 / m = 5 \tmop{mm}
/ \text{s}$. This is demonstrated by considering the ratios $\sigma_{p,
\text{T}} / p_0$ and $\Delta p / p_0$, where $\sigma_{p, \text{T}}$ denotes
the momentum uncertainty exhibited by the longitudinal trap ground state $|
\psi_{\text{T}} \rangle$, reading in the harmonic approximation $\sigma_{p,
\text{T}} = \sqrt{\hbar \omega_{\text{T}} m}$. This yields the ratio
$\sigma_{p, \text{T}} / p_0 = 0.024$, while the pulse duration $T = 60
\tmop{ms}$ implies $\text{$\Delta p / p_0$ = 0.11}$. We can exploit the
smallness of these ratios in order to calculate the normalization $\|
\tilde{C} \|^2$ according to (\ref{SpectrumNorm}). For this, we can focus on a
single-pulse contribution, since in the limit $\tau \gg T$ the spectrum
(\ref{NonNormalizedSpectrum}) represents the superposition of two spatially
distinct dissociation states, which contribute equally to the norm. Within the
range $\sigma_{p, \text{T}}$ where $\langle p_{\tmop{cm}} | \psi_{\text{T}}
\rangle$ is non-negligible, the Feshbach contribution $\text{$\tilde{C}
(p^2_{\tmop{cm}} / 4 m \hbar + p^2_{\tmop{rel}} / m \hbar + 2
\omega_{\text{G}})$}$ to the right hand side of (\ref{NonNormalizedSpectrum})
is only weakly dependent on the center-of-mass momentum $p_{\tmop{cm}}$, which
permits to replace $p_{\tmop{cm}}$ by its mean value $p_{\tmop{cm}} = 0$. The
remaining integral over the relative momentum $p_{\tmop{rel}}$ can be
evaluated in good approximation by linearizing its argument in the vicinity of
$p_0$. Setting $p^2_{\tmop{rel}} - p^2_0 \approx 2 p_0 (p_{\tmop{rel}} - p_0)$
within the region of non-negligible $\tilde{C} (p^2_{\tmop{rel}} / m \hbar + 2
\omega_{\text{G}})$ thus yields
\begin{equation}
  \label{Norm} \| \tilde{C} \|^2 = 2 \pi T^2 \Delta p^2 / p_0 .
\end{equation}
With this, and realizing that the momentum dependent phase factor on the right
hand side of (\ref{NonNormalizedSpectrum}) effects a free time evolution, we
end up with a longitudinal asymptotic wave packet of the form
\begin{equation}
  \label{DTEState} | \Psi_z \rangle = \left( \widehat{\text{U}}_{z,
  \tau}^{(0)} + \mathe^{\mathi \phi_{\tau}} \right) | \Psi_0 \rangle /
  \sqrt{2} .
\end{equation}
Clearly, it is the superposition of an early and a late dissociation
contribution. Note that the additional free time evolution of the early state
component effects a dispersion-induced distortion with respect to the late
state component. The inferred single-pulse momentum spectrum reads
\begin{eqnarray}
  \label{SinglePulseSpectrum} \langle p_{\tmop{cm}}, p_{\tmop{rel}} | \Psi_0
  \rangle &= &\frac{\sqrt{p_0}  \bar{p}^2 \tmop{sinc} \left[ \left(
  p^2_{\tmop{cm}} / 4 + p^2_{\tmop{rel}} - p^2_0 \right) / \Delta p^2
  \right]}{\sqrt{\pi} \Delta p \left( p^2_{\tmop{cm}} / 4 + p^2_{\tmop{rel}} -
  p^2_0 + \bar{p}^2 \right)}
\nonumber\\
 && \times \langle p_{\tmop{cm}}| \psi_{\text{T}} \rangle .
\end{eqnarray}
The state $| \Psi_0 \rangle$ describes a pair of symmetrically
counterpropagating particles, in the sense that each particle can propagate in
both directions, provided the other particle takes the opposite direction. The
corresponding momentum distribution $| \langle p_{\tmop{cm}}, p_{\tmop{rel}} |
\Psi_0 \rangle |^2$ is plotted in Fig. \ref{SinglePulseSpectrumPlot} for our
choice of parameters. As anticipated, the momentum spectrum (\ref{SinglePulseSpectrum})
is sharply peaked at $(p_{\tmop{cm}}, p_{\tmop{rel}}) = (0, \pm p_0)$. This is of
great importance for the viability of the DTE Bell test with material particles, as
was shown in {\cite{Gneiting2008a}}.

The double dissociation pulse sequence generates a two-particle state of the
form (\ref{DTEState}), where due to the additional free time evolution of the
early state component the corresponding wave packets trail each other. The
early and late wave packets are sufficiently spatially separated such that
they can be subjected to the required interferometric transformation. The
double pulse dissociation state (\ref{DTEState}) thus has the structure of a
DTE state as described in Section \ref{Introduction}. The dissociation
probability for the double pulse sequence follows from Eqs.
(\ref{DissociationProbability}) and (\ref{Norm}). For our parameters, we get
$|C_{\tmop{bg}} |^2 \approx 0.04$, which confirms the premised small
dissociation fraction and thus justifies a posteriori the weak coupling
approximation in the derivation of (\ref{ClosedChannelAmplitude}).

\section{\label{DTEBellTest}DTE Bell test}

In this section we demonstrate explicitly that the DTE state (\ref{DTEState}),
when subjected to the interferometric transformation described in Section
\ref{Introduction}, produces an experimentally resolvable fringe pattern that
violates a Bell inequality. In brief, the basic idea behind the
interferometers is to endow the early state components with an additional
phase before recombining them with the late state components. The feasibility
of this procedure requires the early and the late wave packets not to overlap
before arriving at the first, switchable mirror. If this condition is met by
the DTE state (\ref{DTEState}) the joint probability for detecting a pair of
counterpropagating atoms in a particular output port combination labeled by
$\sigma_1$, $\sigma_2$ ($\sigma_i = \pm 1$) follows from
\begin{widetext}
\begin{eqnarray}
  P_{\tmop{dte}} (\sigma_1, \sigma_2 | \ell_1, \ell_2) & = & \frac{1}{4} 
  \left[ 1 + \sigma_1 \sigma_2 \tmop{Re} \left\{ \mathe^{- \mathi \phi_{\tau}}
  \int^{\infty}_{- \infty} \text{d} p_{\tmop{cm}}  \int^{\infty}_{- \infty}
  \text{d} p_{\tmop{rel}} | \langle p_{\tmop{cm}}, p_{\tmop{rel}} |
  \Psi^{(+)}_0 \rangle |^2  \label{DetectionProbability} \right. \right.\\
  &  & \left. \left. \times \exp \left( \mathi \frac{p_{\tmop{cm}} (\ell_1 +
  \ell_2)}{2 \hbar} + \mathi \frac{p_{\tmop{rel}} (\ell_1 - \ell_2)}{\hbar} -
  \mathi \frac{\tau (p^2_{\tmop{cm}} / 4 + p^2_{\tmop{rel}})}{m \hbar} \right)
  \right\} \right], \nonumber
\end{eqnarray}
\end{widetext}
as shown in {\cite{Gneiting2008a}}. The two-particle momentum distribution $|
\langle p_{\tmop{cm}}, p_{\tmop{rel}} | \Psi^{(+)}_0 \rangle |^2$ derives from
the symmetric distribution $| \langle p_{\tmop{cm}}, p_{\tmop{rel}} | \Psi_0
\rangle |^2$ by restricting each particle to propagate into a particular
direction, say, particle 1 into positive direction and particle 2 into
negative direction. Hence we have
\[ | \langle p_{\tmop{cm}}, p_{\tmop{rel}} | \Psi^{(+)}_0 \rangle |^2 = 2
   \theta (p_{\tmop{rel}}) | \langle p_{\tmop{cm}}, p_{\tmop{rel}} | \Psi_0
   \rangle |^2, \]
where the additional factor 2 is imposed by normalization. If the two-particle
momentum distribution $| \langle p_{\tmop{cm}}, p_{\tmop{rel}} | \Psi_0
\rangle |^2$ is sufficiently well-behaved and the path length differences
$\ell_1$, $\ell_2$ are chosen such that the early and late wave packets
overlap in the output ports, the detection probability
(\ref{DetectionProbability}) reveals a fringe pattern under variation of
$\ell_1$ and $\ell_2$. It is remarkable that the quality of this fringe
pattern is only affected by the free time evolution between the early and the
late dissociation, whereas any subsequent time evolution does not have any
effect, despite of the ongoing broadening of the wave packets.

\begin{figure}[tb]
  \resizebox{\columnwidth}{!}{\epsfig{file=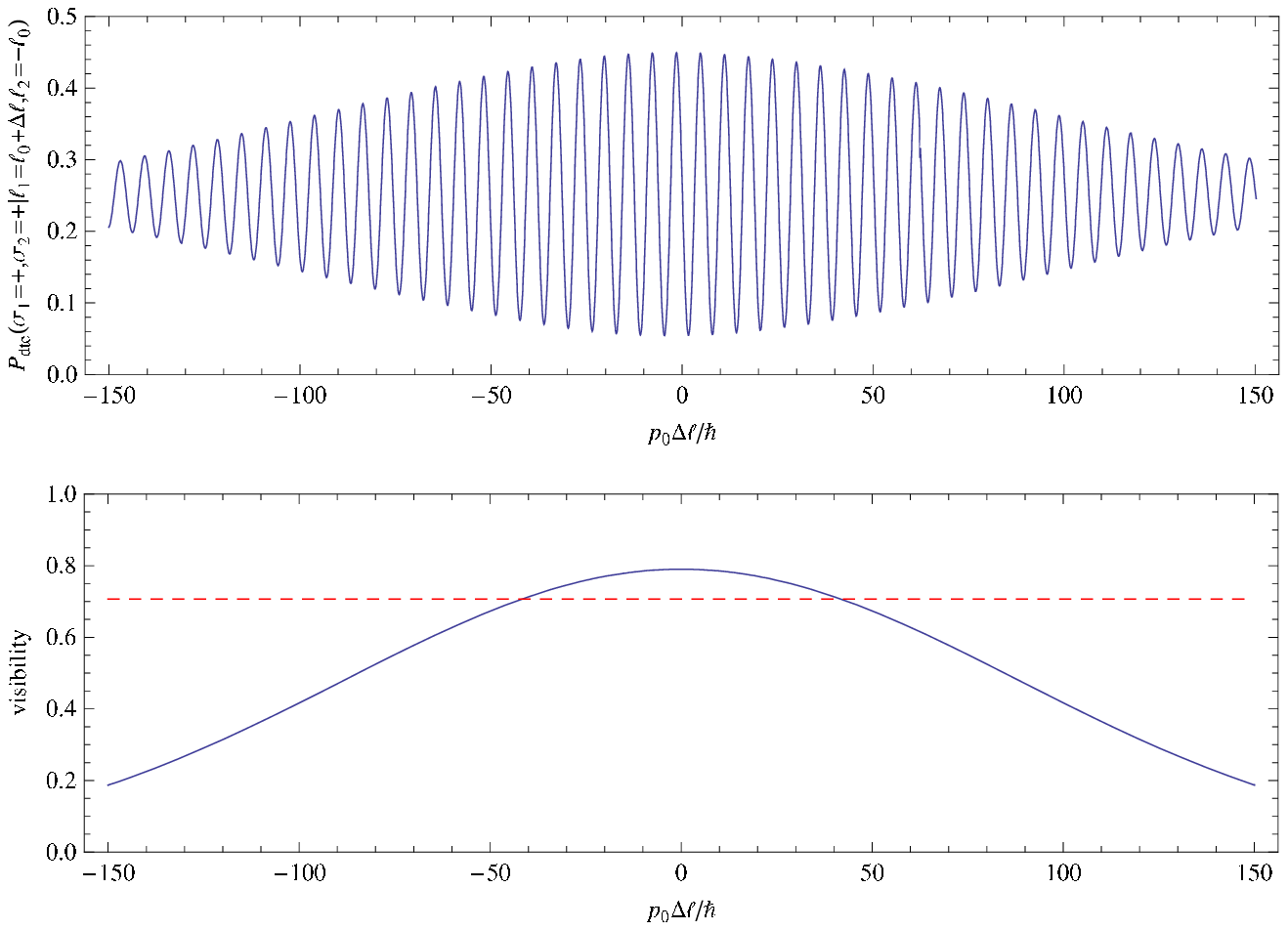}}
  \caption{\label{DTEFringePattern}(color online) Joint probability
  $P_{\tmop{dte}} (\sigma_1, \sigma_2 | \ell_1, \ell_2)$ for detecting a pair
  of counter-propagating atoms in the output port combination $\sigma_1 = + 1$
  and $\sigma_2 = + 1$ under variation of the path length difference $\ell_1$.
  At the outset, $\ell_1$ and $\ell_2$ are chosen such that optimum overlap of
  the early and the late wave packets is achieved in the output ports of the
  beam splitters. The upper plot shows the resulting fringe pattern as it
  would be seen by the experimenter. The characteristic length for its period
  is set by the de Broglie wave length of the relative motion,
  $\lambda_{\tmop{rel}} = 12.4 \mu \text{m}$. The finite range of the
  envelopes reflects the decreasing wave packet overlap with increasing offset
  $\Delta \ell$ from the optimum overlap value. The lower plot highlights the
  corresponding fringe visibility. The dashed line marks the visibility
  threshold that has to be exceeded in order to violate a Bell inequality. A
  sufficient number of fringes exceeds the threshold to manifest the
  corresponding violation. }
\end{figure}

For the single pulse momentum spectrum (\ref{SinglePulseSpectrum}) of the DTE
state (\ref{DTEState}) the joint detection probability
(\ref{DetectionProbability}) can be evaluated numerically. The resulting
fringe pattern for the output port choice $\sigma_1 = + 1$ and $\sigma_2 = +
1$ under variation of $\ell_1$ can be seen in Fig. \ref{DTEFringePattern}.

Specifically, we choose $\ell_1 = \ell_0 + \Delta \ell$ and $\ell_2 = -
\ell_0$, where $\ell_0 = \tau p_0 / m$ defines the optimum overlap path length
difference cancelling the separation between the early and the late wave
packets. The variation of $\ell_2$ or combinations of $\ell_1$ and $\ell_2$
would of course yield similar fringe patterns. The (constant) relative phase
$\phi_{\tau}$ is neglected since it only results in a phase shift of the
fringe pattern, while the upper and lower envelope remain unaffected.

Inspecting Fig. \ref{DTEFringePattern}, we can identify several generic
features of the DTE fringe pattern. First, the characteristic length for its
period is set by the de Broglie wave length of the relative motion
$\lambda_{\tmop{rel}} = h / p_0$. This is natural, since we implement the
relative phase shift by displacing the early and late wave packets with
respect to each other, while the phase variation of a propagating wave packet
is set by its de Broglie wave length. In our case, we get
$\lambda_{\tmop{rel}} = 12.4 \mu \text{m}$. This means that, in order to be
able to detect the fringe pattern, the interferometers have to be kept stable
by about $1 \mu \text{m}$.

The envelope size of the fringe pattern, on the other hand, can be traced back
to the unavoidable offset from the optimum wave packet overlap when varying
the path length differences. One expects no interference if there is no
overlap in the beam splitter output ports, resulting in the uncorrelated
probability distribution of $25\%$ per output port combination. Indeed, the
width of the fringe pattern, which is of about $200 \mu \text{m}$, matches
approximately the initial position spread of the single-particle wave packets.
Minor modifications to this can be traced back to the dispersion-induced
distortion between the early and the late wave packets.

However, the overall visibility reduction, which prevents the fringe pattern
to vary with its maximum amplitude of $0.25$ even in the case of optimum
overlap, is a genuine effect of the dispersion-induced distortion between the
early and the late wave packets. If the early and late wave packets were
identical up to displacement, the visibility should not be affected,
independently of their shape {\cite{Gneiting2008b}}. It is mainly this
criterion which decides whether a single pulse momentum distribution is
sufficiently well-behaved or not. Indeed, the joint detection probability
(\ref{DetectionProbability}) can be evaluated analytically in the case of a
Gaussian momentum distribution, revealing that the overall visibility is
sensitive to the center-of-mass and relative motion momentum spreads
$\sigma_{p, \tmop{cm}}$ and $\sigma_{p, \tmop{rel}}$ {\cite{Gneiting2008a}}.
Only if these are sufficiently small with respect to the relative momentum
$p_0$, the resulting fringe pattern can exceed the visibility threshold $1 /
\sqrt{2}$ as required for violating a Bell inequality. In our case, the
momentum spreads are characterized by $\sigma_{p, \text{T}} / p_0 = 0.024$ and
$(\Delta p / p_0)^2 = 0.012$, yielding a momentum distribution that is
sufficiently well peaked to be capable of violating a Bell inequality, as can
be seen in the lower plot in Fig. \ref{DTEFringePattern}. We emphasize that
the potential violation sustains over a sufficient number of fringes to be
resolvable in experiment.

Another viable approach to attain the expected fringe pattern as determined by
a given (sufficiently well-behaved) momentum distribution is to fit it with a
Gaussian. This permits to apply above mentioned analytic result, as was done
in {\cite{Gneiting2008a}}. However, for spectra of the form
(\ref{SinglePulseSpectrum}) such a least square fit has the tendency to
slightly overestimate the accessible visibility. A more conservative procedure
than least square fitting would define the fit as an upper envelope of the
corresponding momentum distribution. We note that the second momenta of the
momentum distribution resulting from (\ref{SinglePulseSpectrum}) are not
useful for characterizing the widths of the peaks, while the Gaussian fitting
procedure yields a comparatively good approximation in our context.

A final remark should be made on the relative phase $\phi_{\tau}$ between the
early and late state component. As can be seen in
(\ref{DetectionProbability}), it also enters the joint detection probability
and hence influences the fringe pattern. It was neglected in our numerical
investigation since it only effects a phase shift of the fringe pattern. Of
course, this assumes that $\phi_{\tau}$ remains constant from shot to shot,
while an uncontrolled variation of $\phi_{\tau}$ already on the order of $100
\tmop{mrad}$ would spoil the fringe pattern. In our Feshbach calculation
$\phi_{\tau}$ is determined by
\[ \phi_{\tau} \simeq [2 U_{\text{T}} \tau - \mu_{\tmop{res}} \Delta B T +
   \mu_{\tmop{res}} (B_{\tmop{res}} - B_0) \tau] / \hbar + \omega_{\text{G}}
   \tau . \]
Hence, we have to require that the magnetic field pulses can be kept stable
and are reproducible from shot to shot with a relative accuracy of $10^{- 5}$.

\section{\label{Conclusions}Conclusions}

We described an experimental scenario for the generation of a pair of
dissociation-time entangled atoms and the subsequent interferometric Bell
test. Assuming viable experimental conditions, the generated state of two
counterpropagating atoms was shown to be capable of violating a Bell
inequality.

The experiment is based on the Feshbach-induced dissociation of a diatomic
molecule, where the entanglement of the dissociating atoms is achieved by
applying a sequence of two magnetic field pulses. Hence, no complicated
manipulation of the molecular state is required, the only experimental demand
is to keep the molecules trapped in a controlled way and to carry out the
magnetic field pulses with sufficient accuracy. The atom guides and the
required switched, asymmetric Mach-Zehnder interferometers are implemented by
laser beams. Since only single-particle interference has to be performed, the
interferometers on the opposite sides do not have to be kept stable with
respect to each other. This allows to reach macroscopic distances between the
two measurement devices, in our scenario on the order of 10 centimeters. The
procedure is completed by position measurements that only have to be able to
distinguish the chosen beam splitter output port for each atom. This does not
require prominent spatial and temporal resolution and is conveniently
accomplished by resonant laser beams.

The violation of a Bell inequality on the basis of dissociation-time
entanglement would be the manifestation of non-classicality in the
{\tmem{motional}} state of two {\tmem{material}} particles, without resorting
to internal degrees of freedom like spin. Here not only the two atoms are
separated on a scale of several centimeters, but also the early and the late
wave packets are separated by a distance of $5 \tmop{mm}$ on each side. Such a
superposition of two consecutively propagating wave packets for each atom,
which would be established in this experiment, constitutes a macroscopic
violation of the classical particle concept of localization and therefore
presents a worthwhile feature of its own.

\end{document}